\documentclass[aps,twocolumn,showpacs,prl,superscriptaddress]{revtex4-1}
\usepackage{mathtools}
\usepackage{bm}
\usepackage{dsfont,amsthm,amsbsy}
\usepackage{verbatim}
\usepackage{amssymb}
\usepackage{amsmath}
\usepackage{bbm}
\usepackage{graphicx}
\usepackage{epstopdf}
\usepackage{subfigure}
\usepackage{natbib}
\usepackage{epsfig}
\usepackage{amsfonts}
\usepackage{mathrsfs}
\usepackage{sidecap}
\usepackage{lipsum}
\usepackage[toc,page,title,titletoc,header]{appendix}
\usepackage[colorlinks,linkcolor=blue,citecolor=blue,anchorcolor=blue,urlcolor=blue]{hyperref}
\usepackage{hyperref}
\usepackage{resizegather}
\usepackage{tikz}
\usepackage{float}
\usepackage{mathbbol}
\usepackage[normalem]{ulem}
\usepackage{cancel}
\usepackage{upgreek}
\usepackage{relsize}

\newcommand{\bl}{\begin{aligned}}
\newcommand{\el}{\end{aligned}}

\def\be{\begin{equation}}
\def\ee{\end{equation}}

\def\bi{\begin{itemize}}
\def\ei{\end{itemize}}
\def\bn{\begin{enumerate}}
\def\en{\end{enumerate}}
\def\bea{\begin{eqnarray}}
\def\eea{\end{eqnarray}}

\def\ba{\begin{array}}
\def\ea{\end{array}}
\def\bd{\begin{displaymath}}
\def\ed{\end{displaymath}}

\begin{document}
\title
{{Topological Defects from Quantum Reset Dynamics}}


\author{R. Jafari}
\email[]{raadmehr.jafari@gmail.com}
\affiliation{Physics Department and Research Center OPTIMAS, RPTU University Kaiserslautern-Landau, 67663 Kaiserslautern, Germany}

\author{Henrik Johannesson}
\email[]{henrik.johannesson@physics.gu.se}
\affiliation{Department of Physics, University of Gothenburg, SE 412 96 Gothenburg, Sweden}

\author{S. Eggert}
\email[]{eggert@physik.uni-kl.de}
\affiliation{Physics Department and Research Center OPTIMAS, RPTU University Kaiserslautern-Landau, 67663 Kaiserslautern, Germany}


\date{\today}

\begin{abstract}
{\color{black}  We analyze mechanisms for universal out-of-equilibrium dynamics near criticality by \color{black} exploring the effect of randomized quantum resetting (QR) under a finite-time quench across a quantum phase transition. Using the transverse-field Ising chain as a generic {\color{black}model} and exploiting its exact solution,} QR is found to cause a crossover of the scaling of the {\color{black} topological} defect density with the time scale $\tau$ of the quench, from Kibble-Zurek to anti-Kibble-Zurek scaling as $\tau$ increases. This reflects a competition between non-adiabatic quench-driven excitations and QR, giving rise to local minima of the defect densities at optimal annealing times. These times and the corresponding local minima are shown to scale as universal power laws with the rate of QR. {\color{black} Additional results for the scaling of the mean excess energy suggest that a system driven across a quantum critical point exhibits the same scaling behavior under a linear quench with QR as with uncorrelated noise.}  
 \end{abstract}

\maketitle

{\bf {\small Introduction.}} The study of nonequilibrium quantum matter has surged in the last two decades, driven by experimental advances across various platforms such as ultracold gases \cite{Bloch2008}, {\color{black} Rydberg atom arrays} \cite{Browaeys2020}, superconducting circuits \cite{Blais2021}, and a growing class of various quantum materials \cite{Keimer2017}. These developments have raised a number of theoretical challenges, including the quest for universal mechanisms that govern the out-of-equilibrium dynamics of quantum many-body systems \cite{Polkovnikov2011}. 

One of the few established paradigms for universality away from equilibrium is the Kibble-Zurek (KZ) mechanism \cite{Kibble1976,Zurek1985}, extended in Refs.~\cite{Damski2005,Zurek2005,Dziarmaga2005,Dziarmaga2010} to the quantum realm: A system driven into a broken-symmetry phase by varying a control parameter across the critical point of a continuous equilibrium phase transition is predicted to violate adiabaticity due to critical slowing down, leading to the appearance of topological defects. At the core of the KZ mechanism is a universal scaling law $n \sim \tau^{-\beta}$ that governs how the average density of defects {\color{black} $n$} depends on the time scale $\tau = |dt/d\lambda|$ (inverse quench rate) of a linear quench $\lambda(t) = \mbox{const.} \pm t/\tau$ across the critical point, with the exponent $\beta$ determined by the dynamic and correlation length critical exponents of the phase transition. 

A variety of experiments on near-isolated quantum systems show support for the KZ mechanism (see Ref.~\cite{Campo2014} and references therein), however, environmental disturbances can significantly influence the dynamics and cause its breakdown \cite{Griffin2012,Nigmatullin2016,Dario2008,Dario2009,Anirban2016,Puebla2020,Singh2021,Sadeghizade2025}, also when the interaction with the environment is weak. Some of these experiments suggest an {\em increase} of the average defect density at larger time scales, characteristic of anti-KZ behavior \cite{Anirban2016,Singh2021},  
$n \sim \kappa\tau$, with $\kappa$ the rate by which {\color{black} effective noise from the environment} induces excitations in the system. This raises the question about the robustness of KZ scaling. 
Which types of interactions with an environment preserve KZ and which instead yield anti-KZ behavior? 
{\color{black} 
This is an important question for the applicability of KZ scaling as a 
standard testbed for validation of various quantum annealing schemes \cite{Bando2020,Rajak2023} $-$ key to optimization processes on quantum simulation devices \cite{Chandra2010}. 
As we show in this Letter,  
interactions with the environment in the form of so-called {\em quantum resetting} (QR) \cite{Evans2020,Mukherjee2018,Friedman2017,Thiel2018, Yin2023,Rose2018,Gabriele2022,Kulkarni2023,Perfetto2021} 
systematically induce anti-KZ scaling in a quench across a quantum phase transition.  Moreover, the behavior shows data collapse and is
described by universal critical exponents.

QR is {\color{black} typically} realized} by stochastic resets of the driving of a closed system across an equilibrium quantum phase transition \cite{Evans2020,Mukherjee2018,Friedman2017,Thiel2018, Yin2023,Rose2018,Gabriele2022,Kulkarni2023,Perfetto2021}. The time-evolved density matrix in this way gets reset to its initial value at random times. 
QR is a quantum descendant of reset dynamics known to accelerate search processes and induce non-equilibrium steady states in classical systems (see Ref.~\cite{Evans2020} and references therein). The latter property is inherited by QR dynamics where the unitary time evolution of a closed quantum system repeatedly interrupted by resets asymptotically results in a steady-state mixed density matrix, exhibiting nonzero off-diagonal elements \cite{Mukherjee2018}. Additional features uncovered by studies of quantum walks with QR  \cite{Friedman2017,Thiel2018, Yin2023}, open system dynamics resulting from QR \cite{Rose2018,Gabriele2022}, and entanglement generation from QR \cite{Kulkarni2023}, suggest QR as a tool to achieve nonequilibrium steady states with certain specified properties \cite{Perfetto2021}. 

Here, we shall focus on the effect of QR on the generation of topological defects in a transverse-field Ising (TFI) chain when driven by a linear quench across one of its equilibrium quantum critical points. 
{\color{black}
The TFI chain plays a paradigmatic role in the study of quantum phase transitions \cite{SubirBook} and KZ physics \cite{Dziarmaga2005,Dziarmaga2010}  
due to exact solvability \cite{Pfeuty1970} combined with possible
realizations on experimental platforms, {\color{black} which may also allow for QR} \cite{Bao2025,Michel2025}.}

{\bf {\small Preliminaries.}} The Hamiltonian of the TFI chain with a time-dependent magnetic field can be written as
%
\begin{equation}
H(t) =- \sum_{n=1}^{N}( \sigma_{n}^{x} \sigma_{n+1}^{x}
 + h(t) \sigma_{n}^{z}),
\label{eq:Ising}
\end{equation}
%
setting the overall energy scale to unity and choosing $N$ even, with periodic boundary conditions on the Pauli matrices, $\vec{\sigma}_1 \!=\!\vec{\sigma}_{N+1}$. 
We shall take $h(t)\!=\!h_i \pm t/\tau$ to be a linear quench (or ``ramp"), from an initial value $h_i$ at $t\!=\!0$ to a final value $h_f$ at $t\!=\!t_f$. Recall that when $h\!=\!\pm1$, the 
model exhibits quantum phase transitions in the thermodynamic limit, with paramagnetic phases for $|h| \!>\! 1$ and a ferromagnetic 
phase for $|h| \!< \!1$ \cite{Pfeuty1970}. 

{\color{black} The 
instantaneous eigenstates of the TFI Hamiltonian $H(t)$ serve as 
useful basis states, which can be used to analyze the 
non-equilibrium solution  below.
These states are obtained via} a Jordan-Wigner transformation to fermionic creation- and annihilation operators $c_j^{\dagger}$ and $c_j$ \cite{LSM1961} {\color{black} whereby} the transformed TFI Hamiltonian becomes block diagonal, $H(t)\!=\!H_{\text{e}}(t) \otimes 
H_{\text{o}}(t)$, with $H_{\text{e/o}}$ acting on states in the Fock subspace with an even/odd number of fermions. Choosing to work in the even subspace where the fermions obey antiperiodic boundary conditions, $c_{N+1}\!=\!-c_1$, the Fourier transformation $c_j \!=\! N^{-1/2}\sum_k e^{i(jk {\color{black}-\pi/4})}c_k$ turns $H_{\text{e}}(t)$ into a sum over decoupled first-quantized mode Hamiltonians ${\cal H}_k(t)$,  
%
\bea
\label{eq:Nambu}
H_{\text{e}}(t) = \sum_{k>0} C_k^\dagger {\cal H}_{k}(t) C_k. 
\eea
%
Here, $C_k^{\dagger} \!=\! (c_k^{\dagger} \, c_{-k})$ are Nambu spinors and ${\cal H}_{k}(t)\!=\! h_{k}(t)\sigma^z \!+ \!\Delta_k\sigma^x$ where $h_{k}(t)\! =\! 2(h(t)-\cos(k))$ and $\Delta_k = 2\sin(k)$, with $k\!\in \!\{\!\pm \pi/N, \pm 3\pi/N, \ldots, \pm (N-1)\pi/N \}$ as implied by the antiperiodic boundary conditions \cite{Mbeng2024}.    

An instantaneous Bogoliubov transformation $\gamma_k(t)\! =\! \cos\theta_k(t) \,c_k \!-\! \sin{\theta_k(t)}\,c^\dagger_{-k}$ diagonalizes $H_e(t)$, 
%
\begin{equation}
\label{eq:diagonalized}
H_{\text{e}}(t) = \sum_{k} \varepsilon_k(t){\color{black} (\gamma^\dagger_k(t) \gamma_k(t) - \mathsmaller{\frac{1}{2}})},
\end{equation}
%
with $\theta_k(t)\! =\! {\color{black}\arctan(\Delta_k/[h_k(t)\!-\!\varepsilon_k(t)])}$ and $\varepsilon_k(t) \!=\! {\color{black}\sqrt{h_k^2(t) + \Delta_k^2}}$. 
The eigenstates of $H_{\text{e}}(t)$ {\color{black} are} 
%
\begin{eqnarray}
\label{eq:statefactorization}
|\phi_k^-(t)\rangle &=& (\cos\theta_k(t) + \sin\theta_k(t)\, c_k^\dagger c_{-k}^\dagger)|0\rangle \nonumber \\ 
|\phi_k^+(t)\rangle &=& (-\sin\theta_k(t) + \cos\theta_k(t) \,c_k^\dagger c_{-k}^\dagger)|0\rangle,
\end{eqnarray}
%
with energies $\varepsilon^{\mp}_k(t) \!=\! \mp \varepsilon_k(t)$, and where $|0\rangle$ is the vacuum of $\{ c_k, c_k^\dagger \}$. 
By repeating the procedure above for $H_{\text{o}}(t)$, one establishes that the instantaneous ground state 
of the full TFI Hamiltonian $H(t)$ lies in the even subspace for finite $N$, allowing us to initialize the quench in the ground state of the system. 

{\bf {\small Topological defects from quenching without QR.}} 
A topological defect in the TFI chain takes the form of a domain wall (or ``kink") in the ferromagnetic phase $|h|\!<\!1$, being a boundary between regions with different spin orientations \cite{SubirBook}. These defects are delocalized and in 1-1 correspondence with the quasiparticle excitations $\gamma_k^\dagger |\phi_k^-(t)\rangle$ \cite{Mbeng2024}, appearing pairwise in the even subspace and having energies $\varepsilon_k(t)$ according to Eq.~(\ref{eq:diagonalized}). It follows that the expected number of defects ${\cal N}_0$ at the end of a quench, from the paramagnetic into the ferromagnetic phase, is given by ${\cal N}_0 \!= \!\sum_k^N \langle \psi_k(t_f) | \gamma^\dagger_k(t_f) \gamma_k(t_f) | \psi_k(t_f) \rangle$ for large $N$, with $|\psi_k(t)\rangle$ being the state of mode $k$ at time $t$,
%
\begin{equation} \label{eq:timeevolved}
|\psi_k(t)\rangle \!=\! \alpha^+_k(t)|\phi_k^+(t)\rangle + \alpha^-_k(t)|\phi_k^-(t)\rangle.
\end{equation}
%
Note that while the amplitudes $\alpha^{\pm}_k(t)$ in Eq.~(\ref{eq:timeevolved}) are time evolved from the initial state, $|\phi^{\pm}_k(t)\rangle$ are instantaneous. 

{\color{black} It can be verified that} $\gamma_k^\dagger(t)\gamma_k(t) |\phi_{k}^+(t)\rangle\!\!=\!\!|\phi_k^+(t)\rangle$ and $\gamma_k^\dagger(t)\gamma_k(t) |\phi_k^-(t)\rangle \!=\! 0$, so it follows from the above that the expected number of defects is given by
%
\begin{equation}
\label{eq:defectnumber}
{\cal N}_0 = \sum_{k} \langle \phi_k^+(t_f)|\rho_{0,k}(t_f)|\phi^+_k(t_f)\rangle,
\end{equation}
%
where $\rho_{0,k}(t) \!= \! |\psi_k(t)\rangle \langle \psi_k(t)|$ is the density matrix in the $\{ |\phi_k^{\pm}(t)\rangle\}$ basis. By exploiting the equivalence of $|\phi_k^{\pm}(t)\rangle$ to the first-quantized eigenstates of ${\cal H}_k(t)$, realized by the mapping $|0\rangle \rightarrow (0\, 1)^T,\ c_k^\dagger c_{-k}^\dagger|0\rangle \rightarrow (1 \,0)^T$ in Eq.~(\ref{eq:statefactorization}), the von Neumann equation for $\rho_{0,k}(t)$ can be solved exactly using a result by {\color{black} Vitanov and Garraway} for Landau-Zener-type problems \cite{Vitanov1996}. This yields the expected KZ scaling ${\cal N}_0 \!\sim\! \tau^{-1/2}$, first derived for the TFI chain by Dziarmaga via a different route {\cite{Dziarmaga2005}. 

{\bf {\small Topological defects from quenching with QR.}}} 
When quenching with QR, the unitary time evolution gets interrupted at random times by resets to the initial state at time $t\!=\!0$. We shall focus on ideal resets where the projection to the initial state has unit probability. Moreover, as is standard, the distribution of random resets will be taken as Poissonian. For each mode $k$, these conditions are enforced by the QR protocol for infinitesimal time evolution 
%
\bea
|\psi_{k}(t\!+\!dt)\rangle \!=\!
\begin{cases} |\psi_k(0)\rangle, \ \ {\cal P} = r dt \\
\\ 
(1-i {\mathcal H}_k(t)\, dt)\, |\psi_{k}(t)\rangle, \ \ {\cal P} =1\!-\!r dt, 
\end{cases}
\label{eq:resetSchrodinger}
\eea
%
where ${\cal P}$ denotes the probability and $r$ is the rate of resets. Hence, within a time interval $dt$, mode $k$ is reset to its initial state with
probability $rdt$, or, with the complementary probability $(1-r\, dt)$, the mode is time-evolved by ${\mathcal H}_{k}(t)$. With ${\cal H}_k(t)$ being the first-quantized Hamiltonian, note that the states in the protocol 
are the first-quantized equivalents of the states in Eq.~(\ref{eq:timeevolved}), obtained, as above, from the mapping $|0\rangle \rightarrow (0\, 1)^T,\ c_k^\dagger c_{-k}^\dagger|0\rangle \rightarrow (1 \,0)^T$ in Eq.~(\ref{eq:statefactorization}).  
%
\begin{figure*}[t]
\centerline{
\includegraphics[width=0.33\linewidth]{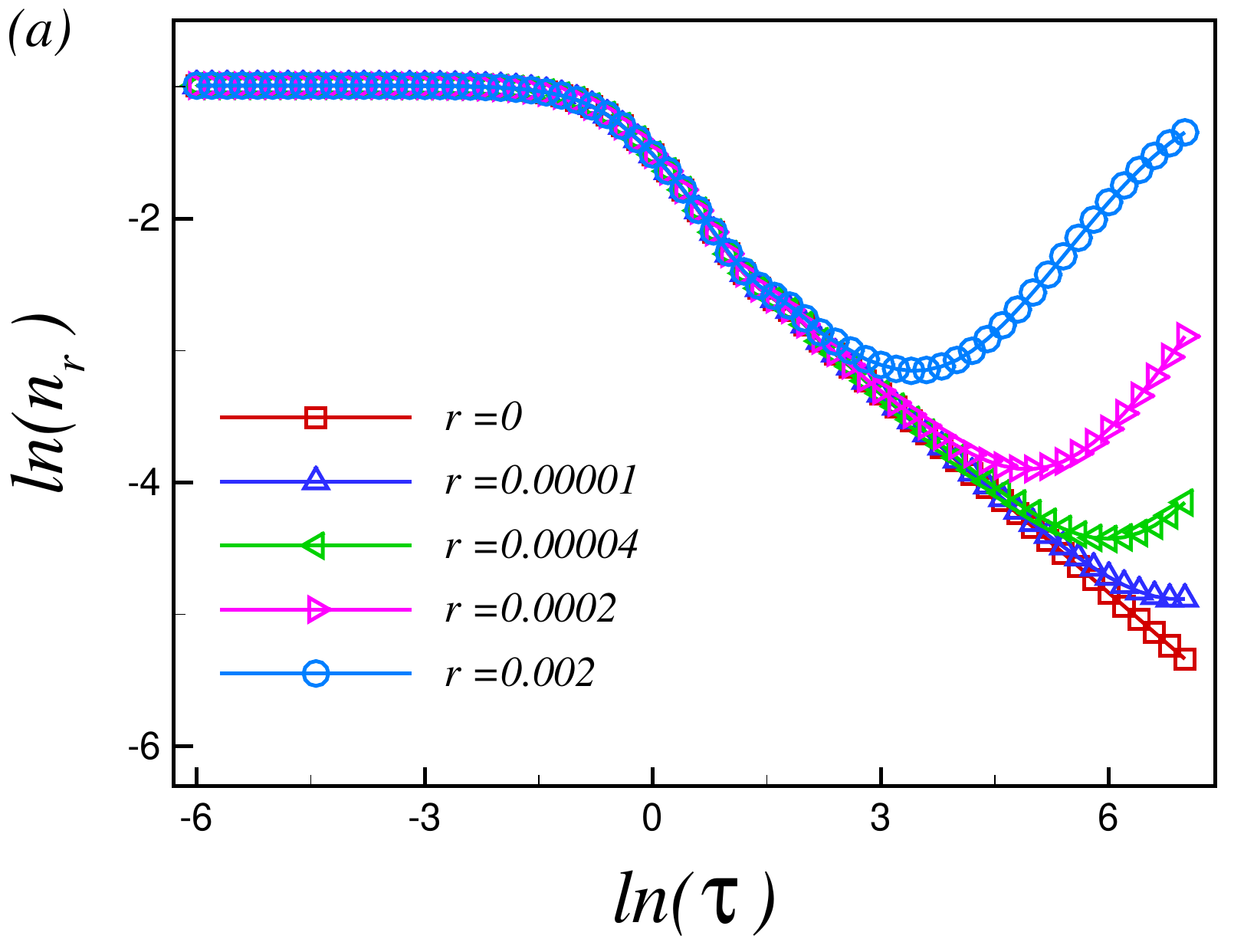}
\includegraphics[width=0.33\linewidth]{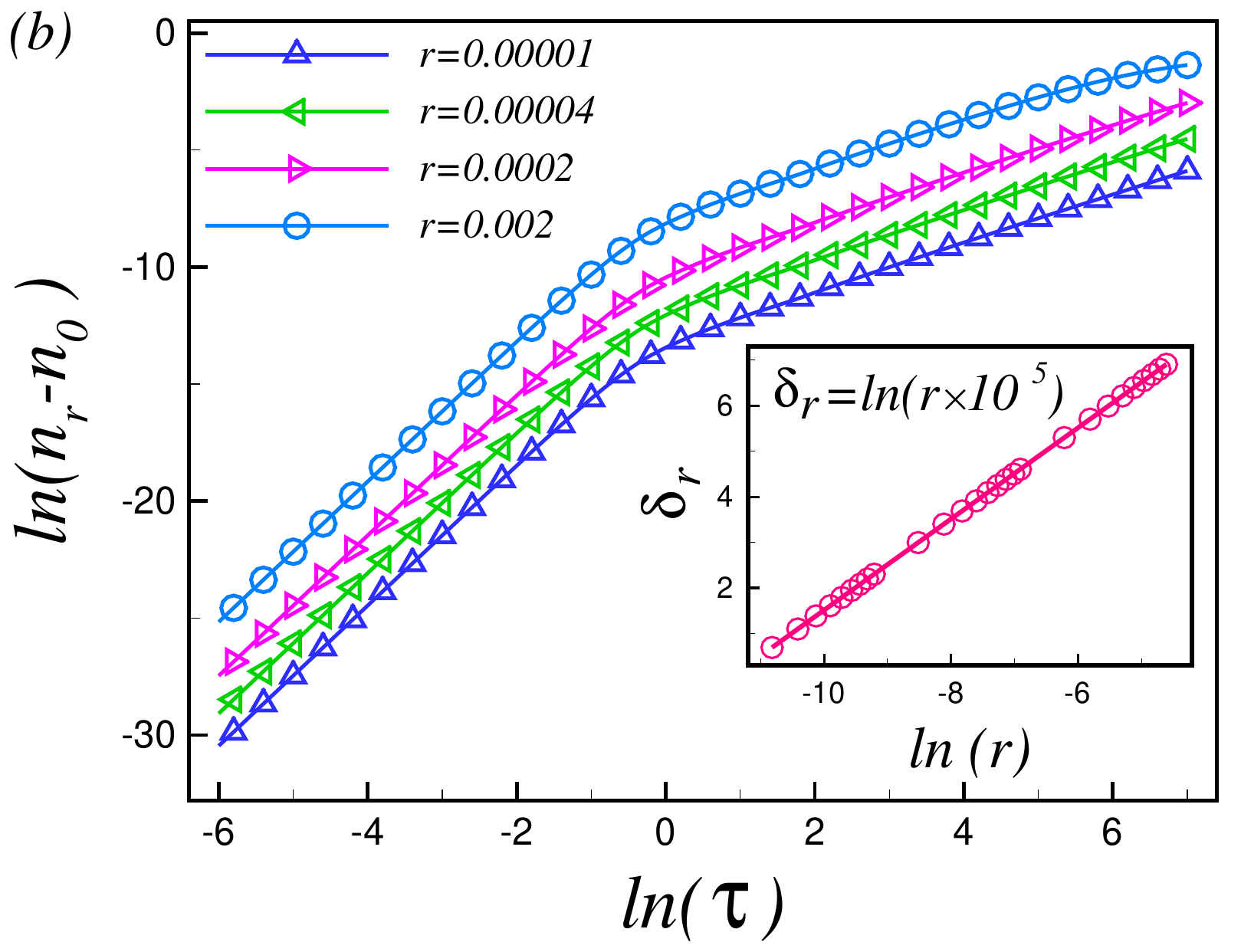}
\includegraphics[width=0.33\linewidth]{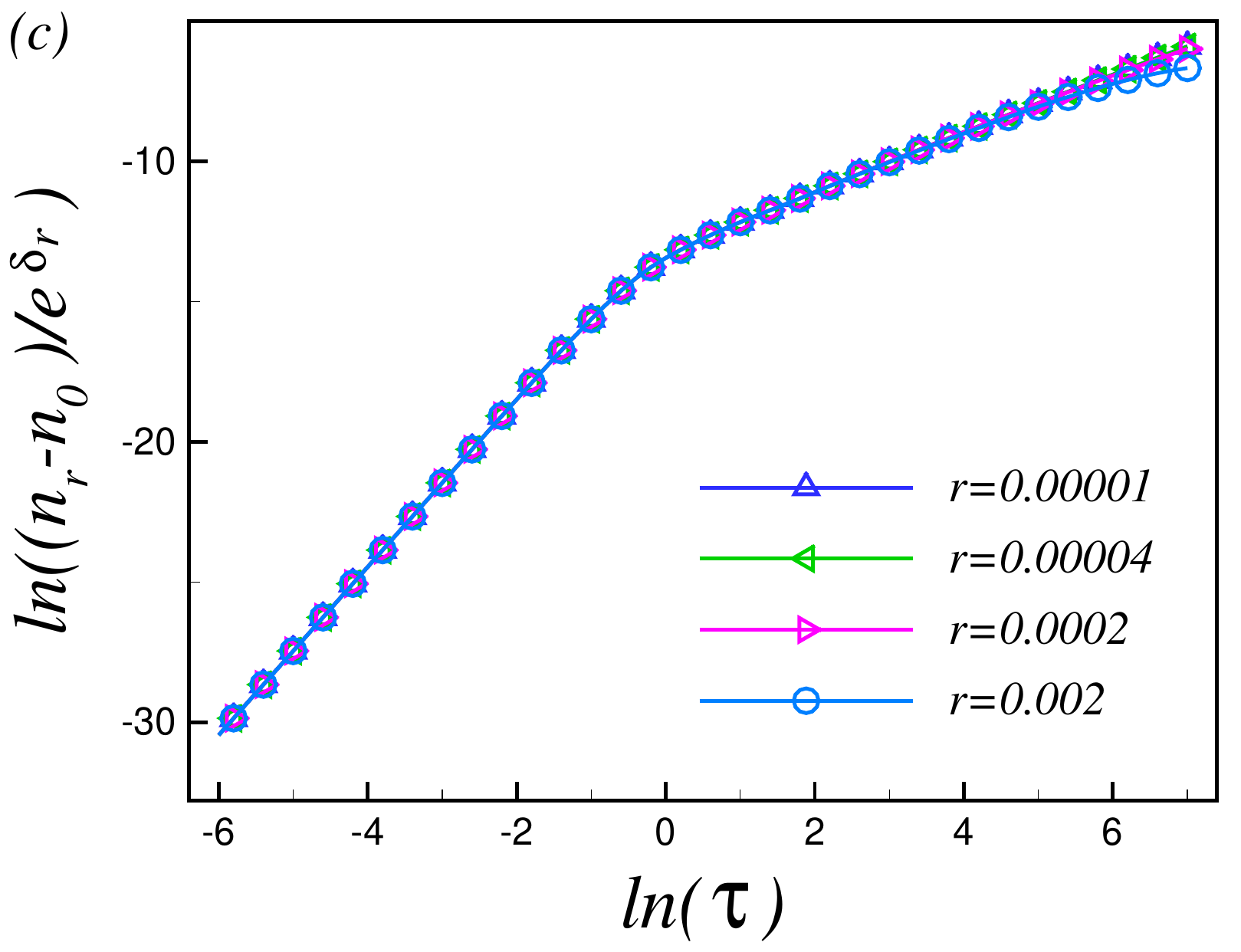}}
\caption{ (a) Log-log plot of the density of defects $n_r$ generated by a quench with QR from $h_i(0) = 2$ to $h_f(2\tau) = 0$ as a function
of the quench time scale $\tau$ for different values of QR rate $r$. (b) Difference $\delta n_r = n_r-n_0$ between the density of generated defects 
from the quench in (a) {\em with} [$n_r$] and {\em without} [$n_0$] QR. (c) Data collapse of the curves in panel (b) after rescaling $\delta n_r \rightarrow e^{-\delta_r}\delta n_r$. 
All data in the figure are obtained from a chain with 1000 sites.}
\label{fig1}
\end{figure*}

Guided by Eq.~(\ref{eq:defectnumber}) for a quench { without} QR, the problem to determine the expected number of defects {\em with} QR boils down to computing the corresponding density matrix. 
While the nonunitary time evolution for a single realization of the QR protocol in Eq.~(\ref{eq:resetSchrodinger}), call it $\eta$, is deterministic, the process introduces an element of classical randomness in the sense that it varies from one realization to another. The mixed density matrix $\rho_{r,k}(t)$ in the presence of QR is therefore obtained as the expectation value $\rho_{r,k}(t) = \langle \rho_{\eta,k}(t)\rangle_r$ by averaging over all QR realizations $\{\eta\}$ with rate $r$. As a result \cite{Mukherjee2018},
%
\bea
\rho_{r,k}(t) &=& r \int_0^{t} e^{-r t'}\, \rho_{0,k}(t')\, d t' + e^{-r t}\, \rho_{0,k}(t), 
\label{eq:QRdensitymatrix}
\eea
%
where the last term corresponds to the event that there is no QR within $t \!\in \! [0,t]$. 

By substituting $\rho_{r,k}(t_f)$ for $\rho_{0,k}(t_f)$ in Eq.~(\ref{eq:defectnumber}), we can numerically obtain the expected number of defects ${\cal N}_r$ in a quench with QR of rate $r$.
For definiteness we shall assume that the system is initially [$t\!=\!0$] prepared in its paramagnetic ground state $|\psi(0)\rangle = \otimes_{k} |\psi_k^-(0)\rangle$ with $h_i = 2$ and $|\psi_k(0)\rangle =  |\phi_k^-(0)\rangle$. The magnetic field is then ramped through the quantum critical point $h = 1$ to the final value $h_f=0$ in the ferromagnetic region at $t_f=2\tau$. 

{\bf \small {Results.}} 
The dependence of the expected density of defects $n_r = {\cal N}_r/N$ on the time scale $\tau$ of the quench is shown in Fig.~\ref{fig1}(a) for several values of the
QR rate $r$. For $r\!=\!0$ (no QR), we numerically verify the known KZ scaling law for the TFI chain, $n_0\! \sim\! \tau^{-1/2}$ \cite{Dziarmaga2005} when $\tau \!\gtrsim \!1$.
The plateau that develops for smaller values of $\tau$ reflects the saturation of defects for very fast ramps where KZ scaling breaks down \cite{Zeng2023}.  
Given the scale in Fig.~\ref{fig1}(a), the effect of QR ($r\! \neq \!0$) is invisible on the plateau as well as in the adjacent $r$-dependent $\tau$-intervals where KZ scaling still holds. 
Beyond these intervals, for larger values of $\tau$, the defect densities $n_r$ start to deviate from KZ scaling. This is indicative of a crossover to an anti-KZ regime \cite{Anirban2016,Singh2021} where slower ramps boost the defect density. 
As $\tau$ increases further, $n_r$ passes through a local minimum and then shoots up, characteristic of anti-KZ. Thus, the defect generation appears to be controlled by two competing processes: (i) non-adiabatic quasiparticle excitations, suppressed by increasing $\tau$ as in a KZ scenario \cite{Dziarmaga2005}, and (ii) QR-induced quasiparticle excitations, amplified by increasing $\tau$ as in an anti-KZ scenario \cite{Anirban2016,Singh2021}. Likewise, smaller values of $\tau$ lead to higher non-adiabaticity and provide less opportunity for QR to prevail. As manifest in Fig.~\ref{fig1}(a), the competition between the two processes yield local minima of $n_r$ at time scales $\tau_{\text{opt},r}$, acting as optimal times in a putative quantum annealing scheme \cite{Chandra2010}, to be discussed below.  

Fig.~\ref{fig1}(b) displays scaling of the defect density from QR {\em only}, obtained by subtracting $n_0$ from $n_r$. Extracting numbers from the data, 
QR-induced defects are found to exhibit anti-Kibble-Zurek scaling with a linear growth with $\tau$, $\delta n_r \sim \tau^{\alpha}$ where $\alpha \!= \!1.00 \pm 0.02$ when $\tau > 1$. 
{\color{black}In contrast,} when $\tau < 1$, the growth of defects with $\tau$ scales nonlinearly with an exponent $\alpha^\prime \!=\! 3.000 \pm 0.003$.
The sharp jump of the exponent at $\tau \!=\! 1$ $-$ which happens before the density-saturated plateau has fully developed $-$ is suggestive of some type of criticality in the QR dynamics. As transpires from Fig.~\ref{fig1}(b),
the amplitude of defect growth with $\tau$ {\color{black} appears to be} too small in this region to make the effect visible in an experiment,
{\color{black}but from a theoretical point of view the universal nature of QR dynamics point to an interesting underlying mechanism in that region.}

As seen in Fig.~\ref{fig1}(b), the $\delta n_r$ curves undergo displacements as the value of $r$ varies. An analysis reveals that the shift for a given value of $r$, call it $\delta_r$, 
scales logarithmically with $r$ as $\delta_r\!=\!\ln(r\times10^5)$ (inset of Fig.~\ref{fig1}(b)). This finding implies that the scaling of $e^{-\delta_r}\delta n_r$ with $\tau$ for different 
values of $r$ is universal. Fig.~\ref{fig1}(c) illustrates the universality, showing that all curves from Fig.~\ref{fig1}(b) collapse onto a single graph after the rescaling. 

{\color{black} {\color{black} Summing up the data in} Fig.~\ref{fig1}(a), {\color{black} with input} from Fig.~\ref{fig1}(b), the defect density $n_r$ behaves as} 
%
\begin{eqnarray}
n_r & \!\approx\! &
\begin{cases} {\color{black} h(r)} \tau^{\alpha^\prime} + \Lambda(\tau), \ \ \tau < 1 \\
\\ 
{\color{black} h(r)} \tau^{\alpha} + b\tau^{-\beta}, \ \ \tau > 1 
\end{cases} 
\label{eq:fulldata}
\end{eqnarray}
%
with $\alpha^\prime \!=\! 3.000 \pm 0.003, \alpha \!=\! 1.00 \pm 0.02, \beta \!=\! 0.50 \pm {\color{black} 0.01}$. {\color{black} The amplitude $h(r)$ is the $r$-dependent rate} at which QR generates quasiparticle excitations, with $\Lambda(\tau)$ a function that describes the crossover to the density-saturated plateau in Fig.~\ref{fig1}(a) as $\tau$ decreases below unity. {\color{black} Here, $b\!=\! c^{-1/2}$, where $c$ is the speed of quasiparticle excitations  \cite{Zurek2005}, of order unity for $\hbar \!= \!a \!= \!1$, with $a$ the lattice spacing of the TFI chain. Apart from the QR-dependent prefactor $h(r)$}, and the specific error estimates of the exponents, the behavior of $n_r$ for $\tau>1$ is the same as that obtained by adding uncorrelated (white) noise \cite{Anirban2016} or correlated (colored) fast noise \cite{Singh2021,Sadeghizade2025} to a ramp across criticality of the TFI chain. The identity of scaling exponents for the two types of quench dynamics is most intriguing, specifically since the probability distributions in their quench protocols, Poissonian for QR and Gaussian for noise, are different. 
{\color{black} This invites the question if this similarity also extends to other observables, such as the time $\tau_{\text{opt},r}$  of minimal defects 
introduced above. } 
%
\begin{figure}
\centerline{
\includegraphics[width=0.65\columnwidth]{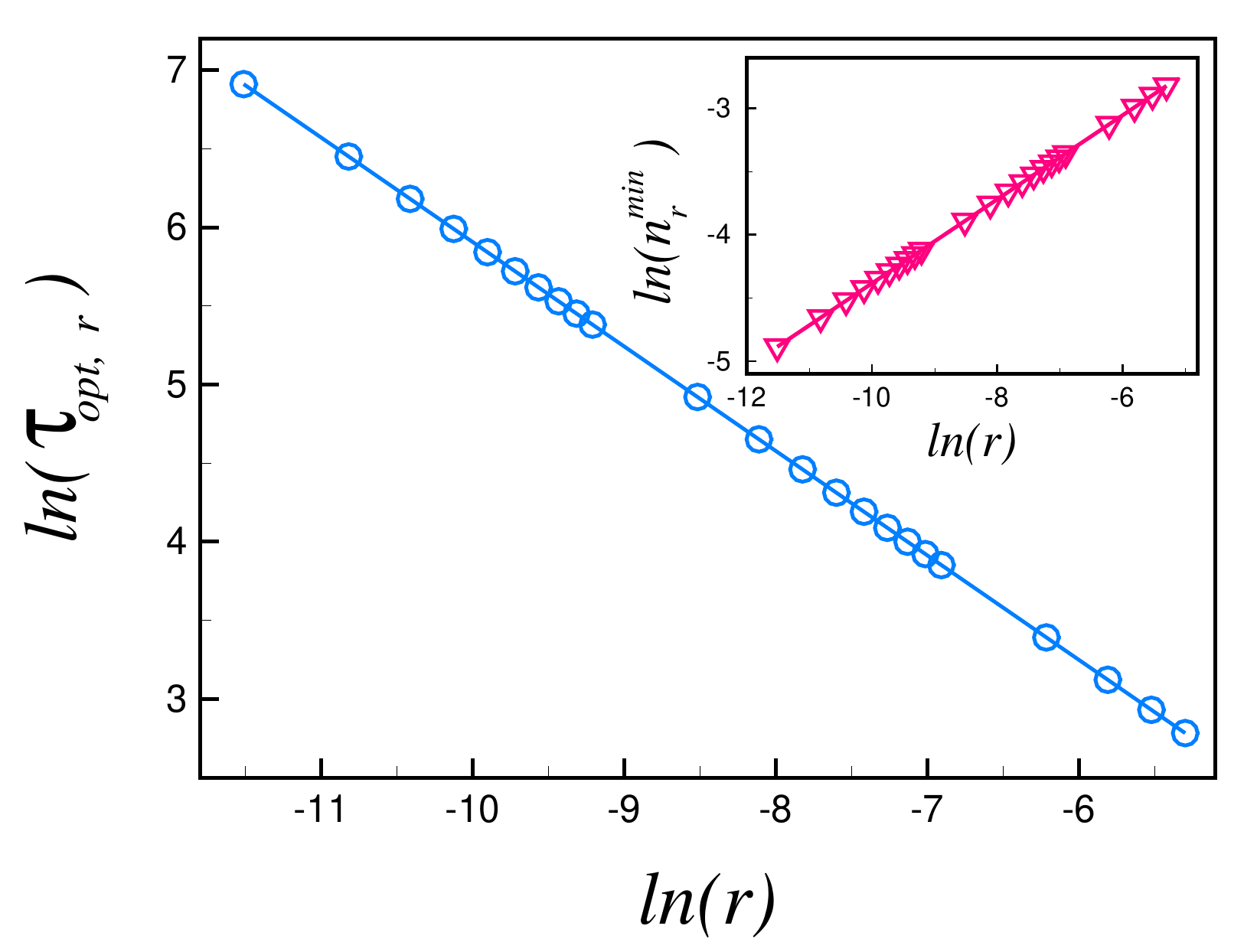}}
\caption{ Scaling of the optimal quantum annealing times $\tau_{\text{opt},r}$ with QR rate $r$. 
Inset: Scaling of the local minima of defect densities $n_r^{\text{min}}$ with $r$.}
\label{fig2}
\end{figure}
%
%
\begin{figure}[t]
\centerline{
\includegraphics[width=0.64\columnwidth]{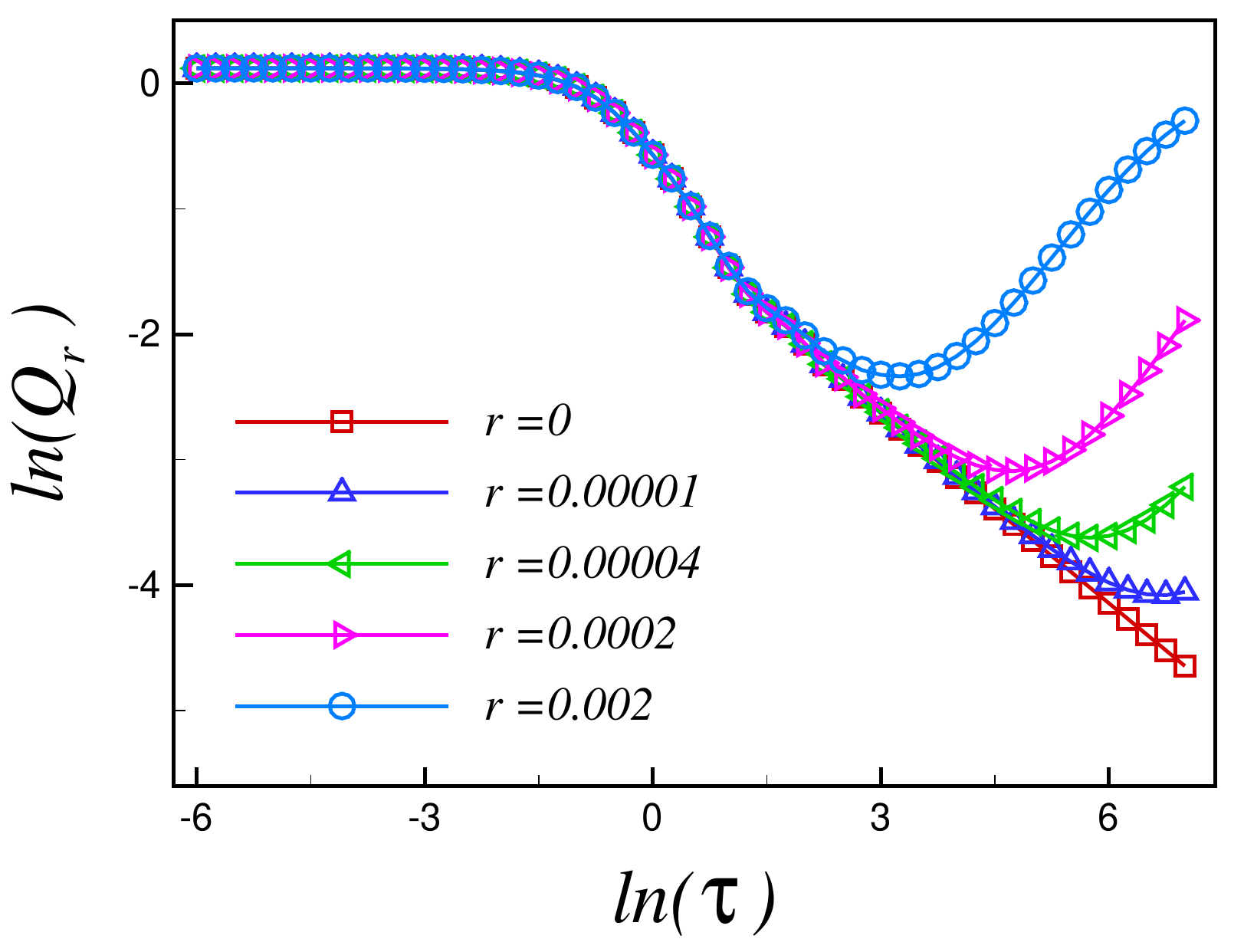}}
\caption{ Residual mean energy $Q_r$ after a quench with QR as a function of the time scale $\tau$ of the quench
for different values of QR rate $r$.}
\label{fig3}
\end{figure}

Optimal times for quantum annealing define the best trade-offs between maximal adiabaticity (slow quench) and minimal decoherence (fast quench), at the same time constraining the defect production when crossing a critical point \cite{Chandra2010}. In the presence of QR, the location of the local minima of the $r$-dependent graphs in Fig.~\ref{fig1} pinpoint distinct optimal times $\tau_{\text{opt},r}$ beyond which there is a fast increase of defects. As follows from the data shown in Fig.~\ref{fig2}, $\tau_{\text{opt},r} \sim r^{-\gamma}$ with $\gamma\!=\!0.664\pm0.002 \simeq 2/3$. Also this scaling exhibits a perfect match between QR and noisy ramp dynamics: The exponent $\gamma$ is the very same as that which governs the scaling of the optimal quench time with noise intensity in a standard quantum annealing protocol with white \cite{Anirban2016} or fast colored \cite{Singh2021,Sadeghizade2025} noise. Our numerical results further show that the local minima of the defect densities $n_{r}^{\text{min}}$ scale with the QR rate as $n_{r}^{\text{min}} \sim r^{0.332 \pm 0.002} \simeq 1/3 \!=\!\gamma/2$ (inset of Fig.~\ref{fig2}).   

The fact that the scaling of defect densities {\em and} optimal annealing times in QR dynamics are identical with those in ramp dynamics with fast noise (uncorrelated or correlated) makes it compelling to analyze additional quantities. One such object, computed also in Ref.~\cite{Anirban2016}, is the mean excess energy of mode $k$ with respect to the corresponding ground state at the end of the ramp \cite{Grandi2010}, 
%
\begin{equation} \label{eq:Q}
Q_r\!=\!\frac{1}{N}\left(\mbox{Tr}(\rho_{{\text r},k}(t_f){\mathcal H}_{k}(t_f))\!-\!\langle\phi_{k}^{-}(t_f)|{\mathcal H}_{k}(t_f)|\phi_{k}^{-}(t_f)\rangle\right)\!.
\end{equation}
%
As suggested by a comparison between 
Fig.~\ref{fig1}(a)
and 
Fig.~\ref{fig3}, one finds that $Q_r$ exhibits the same universal scaling as the defect density in Eq.~(\ref{eq:fulldata}). This result agrees with the noisy ramp dynamics in Ref.~\cite{Anirban2016} which shows identical scaling behaviors for defect densities and excess energies. The universality between the dynamics enforced by the two quench protocols, one with QR and the other with noise, extends to the scaling of the variance of the mean excess energies, as well as to the optimal annealing times extracted from the local minima of the excess energies for different rates $r$ (cf.~Figs.~\ref{fig3}). 

{\small {\bf Nontopological defects from quenching with QR.}} 
Fig.~\ref{fig1}(a) suggests that the KZ mechanism dominates the scaling of the topological defect density in the interval $1 \!< \!\tau \!< \!\tau_{\text{opt},r}$, within which increasing time scales imply a decreasing density. 
Recall that the KZ mechanism rests on the assumption that the critical slowing down in the vicinity of a critical point enforces local selections of a spontaneously broken symmetry, in this way producing topological defects that separate regions with different order \cite{Campo2014}. However, in our numerics we work in the even-parity sector of the fermionized TFI chain with a unique ground state, hence with no spontaneous symmetry breaking in the ferromagnetic region. {\color{black} Still, the KZ exponent $\beta = 1/2$ known for the TFI chain is well reproduced by our data. This is so, since an excitation in the fermionized TFI chain signals the presence of a region of {\color{black}mismatch} between spins {\color{black}$-$ costly in energy $-$} independent of the underlying type of order or its origin.

This becomes manifest when quenching from the ferromagnetic to the paramagnetic phase of the TFI chain where the defects are nontopological, realized by single spin flips \cite{SubirBook}. Here, KZ scaling applies also after a quench from $h_i\!=\!0$ (ferromagnet) to $h_f \!\gg \!1$ (paramagnet) with no spontaneous symmetry breaking in the paramagnetic phase \cite{Dziarmaga2005}. It follows that KZ {\em scaling} has a wider applicability than predicted by the standard KZ {\em mechanism} (based on spontaneous symmetry breaking and an impulse approximation scenario \cite{Campo2014}).} 

{\color{black} To close this section we consider QR in a quench from a ferromagnet to a paramagnet, using
the same} QR protocol as in Eq.~(\ref{eq:resetSchrodinger}) but with $h_i=0$ at $t=0$ and $h_f=2$ at $t_f=2\tau$. All scaling exponents 
$\alpha, \alpha^\prime, \beta$, and $\gamma$ reported above are well reproduced, with roughly the same numerical errors. The only change 
is that the optimal annealing times $\tau_{\text{opt},r}$, signaling the onset of the anti-KZ regimes, {\color{black} are shifted {\color{black} by a small amount compared to Fig.~\ref{fig1}(a).}}

{\small {\bf Summary.}} {\color{black} The universal behavior of quenches through a quantum phase transition 
has been analyzed in the presence 
of interactions with the environment in the powerful form of QR.  The out-of-equilibrium dynamics near criticality with QR clearly
shows {\color{black} the presence of anti-KZ behavior, notably with data collapse and universal exponents}. In particular, 
the interplay of the quench with QR $-$ both processes known to yield nonequilibrium states $-$ is found to produce an intriguing behavior of the defect density. The results reveal that QR increases the number of topological defects as the time scale $\tau$ increases, with a critical value 
at which the defect density abruptly changes its scaling with $\tau$. We also find that the competition between adiabaticity and QR 
yields an optimal annealing time for which the defect density is at a minimum. This optimal time exhibits universal scaling with the QR rate, with the very same exponent as for scaling in the presence of noise in previous studies \cite{Anirban2016,Singh2021,Sadeghizade2025}.

Hence, a system driven across a quantum critical point with Poissonian distributed QR remarkably appears to show the same anti-KZ
scaling behavior as a quench with uncorrelated Gaussian noise, indicating that universal mechanisms are in place for
out-of-equilibrium dynamics near criticality.  This invites further studies 
if this unexpected correspondence holds {\color{black} more generally}, possibly extended to correlated noise and resets.} \\ 
 
{\small {\bf Acknowledgements.}}
SE is thankful for support from the Deutsche Forschungsgemeinschaft via Project A5 in the SFB/TR185 OSCAR. \\

{\small {\bf Data Availability.}} 
{\color{black} The data that support the findings of this article are not publicly available. The data are available from the authors upon reasonable request.}

\bibliography{QR_References}

\end{document}